\documentclass[10pt,aps,prd,twocolumn,showpacs,preprintnumbers,amsmath,amssymb,nofootinbib]{revtex4-1}

\usepackage{epsfig}

\usepackage{amsmath}   
\usepackage{amssymb}
\usepackage{hyperref}  
\usepackage{nameref}   

\usepackage{comment}
\usepackage[table]{xcolor}
\usepackage{graphicx}
\usepackage{booktabs}
\usepackage{dcolumn}
\usepackage{bm}

\usepackage{hyperref}
\hypersetup{colorlinks=true,linkcolor=magenta,anchorcolor=green,citecolor=cyan,filecolor=black,menucolor=black,urlcolor=brown}
\usepackage{slashed}

\newcommand{\be}{\begin{equation}}
\newcommand{\ee}{\end{equation}}

\newcommand{\ba}{\begin{eqnarray}}
\newcommand{\ea}{\end{eqnarray}}

\newcommand{\bea}{\begin{eqnarray}}
\newcommand{\eea}{\end{eqnarray}}

\begin{document}

\title{Constraining self-interacting scalar field dark matter with strong gravitational lensing in cluster scale?}

\author{Raquel Galazo Garc\'ia}
\email{rgalazo@comillas.edu}
\affiliation{Universidad Pontificia Comillas-ICADE, Alberto Aguilera 23, 28015 Madrid, Spain}
\affiliation{Aix--Marseille Universit\'{e}, CNRS, CNES, Laboratoire d'Astrophysique de Marseille, France}
\author{Eric Jullo}
\affiliation{Aix--Marseille Universit\'{e}, CNRS, CNES, Laboratoire d'Astrophysique de Marseille, France}
\author{Emmanuel Nezri}
\affiliation{Aix--Marseille Universit\'{e}, CNRS, CNES, Laboratoire d'Astrophysique de Marseille, France}
\author{Marceau Limousin}
\affiliation{Aix--Marseille Universit\'{e}, CNRS, CNES, Laboratoire d'Astrophysique de Marseille, France}

\begin{abstract}

We study solitonic cores in the Thomas--Fermi regime of self-interacting scalar field dark matter (SI-SFDM) and assess whether strong gravitational lensing by galaxy clusters can detect or constrain them. We build composite halo profiles that join a solitonic core to an outer NFW profile at fixed total mass, and compute their lensing signatures---the deflection angle and the excess surface mass density---for two halos of typical cluster scale, $M_{200}=2\cdot10^{15}$ and $2\cdot10^{14}\,\rm M_\odot$. The soliton radius is fixed by the microphysics ($m$, $\lambda_4$), while its mass is not fixed by these parameters but is constrained by the halo mass budget. Lensing constrains only the radius, and hence the combination $\lambda_4/m^4$; combined with the independent bound on the self-interaction cross section, this places an indirect upper bound on the scalar field mass, $m\lesssim\mathrm{few}\times10^{-5}$~eV, and correspondingly $\lambda_4\sim10^{-18}$. Taking the NFW profile as reference, we quantify the deflection-angle difference produced by the soliton and confront our predictions with the cluster A2390 and with the stacked DES weak-lensing profile to assess when an SI-SFDM core can be distinguished from NFW.

\end{abstract}

\date{\today}

\maketitle

{\noindent\small\textit{Published in Phys.\ Rev.\ D \textbf{114}, 103520 (2026), \href{https://doi.org/10.1103/651s-h6v3}{DOI: 10.1103/651s-h6v3}. \copyright~2026 American Physical Society.}\par}

\section{Introduction}\label{sec:intro}

The cold dark matter (CDM) model is the leading framework for explaining dark matter (DM) in the Universe, with weakly interacting massive particles (WIMPs) being the leading candidates due to strong theoretical and experimental motivation \citep{Jungman:1995,Drees:2004jm,STEIGMAN1985375}. However, despite extensive searches, no direct detection of WIMPs has been achieved \citep{2019JPhG...46j3003S,2014arXiv1411.1925C,2018EPJC...78..203A}. As observational techniques and numerical simulations have advanced, discrepancies at small cosmic scales—commonly referred to as small-scale tensions—have emerged \citep{Weinberg2015,Popolo2017,Nakama2017}. These unresolved tensions \citep{Luzio2020} suggest the need for new physics beyond the standard CDM paradigm to more accurately describe the Universe’s structure \citep{Weinberg2015}.  

In light of these challenges, alternative DM models propose that dark matter could be described by a scalar field (SFDM) with masses ranging from \(10^{-22}\) eV to 1 eV \citep{Hu2000,Hui2017,Goodman:2000tg}. This mass range corresponds to a de Broglie wavelength that can influence structures at galactic and sub-galactic scales, potentially addressing the small-scale tensions observed in the CDM framework.  

Hence, at scales much smaller than the de Broglie wavelength, \(\lambda_{dB} \sim 1/(m v)\), the scalar field exhibits wave-like behavior. These wave effects arise from an additional pressure term in the equation of motion, known as quantum pressure, which originates from the gradient of the scalar field. When quantum pressure counterbalances gravity, it leads to equilibrium configurations known as solitons \citep{Lee:1991ax, 2015PhRvD..92j3513G, 2009PhRvL.103k1301S}. These solitons form at the center of dark matter halos, resulting in a flattened radial density profile at their core. The SFDM model, where the scalar mass serves as the only free parameter, is widely referred to in the literature as Fuzzy Dark Matter or ultra-light dark matter \citep{Hui2017,Hu2000,Schive2014a,Schive2014b,2022PhRvD.105l3528G}.  

When self-interactions are introduced into the SFDM framework (SI-SFDM) \citep{Chavanis15c,Brax:2019fzb}, equilibrium arises from a balance between self-interaction, quantum pressure, and gravity. In the Thomas-Fermi regime \citep{Thomas_1927,Chavanis:2011zi,Chavanis:2017loo,Dawoodbhoy:2021beb,Shapiro:2021hjp}, repulsive self-interactions dominate over quantum pressure, leading to equilibrium configurations governed solely by self-gravity and an effective pressure term. This results in the formation of smooth, finite-size solitons, whose size depends on the mass and self-interaction properties of the scalar field but is independent of the soliton mass. The cosmological evolution of such models has been investigated in various studies \citep{Chavanis17c,Li2014,Nori:2018hud}. Additionally, the formation and evolution of solitons within scalar halos featuring different self-interactions have been examined \citep{PhysRevD.109.043516, PhysRevD.111.063511}, along with the formation and dynamics of vortex lines in rotating scalar dark matter halos, particularly in models with quartic repulsive self-interactions \citep{Brax:2025uaw,Brax:2025vdh}.

Detecting dark matter often relies on studying its gravitational effects. One approach involves studying its influence in the vicinity of black holes, either by analyzing gravitational waves \citep{Boudon2023} or by estimating properties of the black hole environment through observational methods \citep{Bar_2019, Chakrabarti_2022, gomez2024constraining}. Another promising technique for identifying evidence of SI-SFDM, as explored in this work, is the use of strong gravitational lensing (SL). SL provides a powerful tool to constrain the central projected density distribution of dark matter. Studies employing SL have uncovered evidence of cored dark matter distributions in cluster cores \citep{sand04, Newman_2013, Limousin22}. In contrast, other analyses have found steep dark matter distributions in cluster cores \citep{Limousin_2008}, highlighting significant variation across the cluster population. 

SL offers arcsecond-level precision (equivalent to a few kiloparsecs) in constraining the peak of the dark matter distribution. In merging systems, this precision is used to search for offsets between the dark matter and the associated light, a signature expected in self-interacting dark matter (SIDM) models; the absence of such offsets to date places upper limits on the DM self-interaction cross-section \citep{Sirks_2024, Randall_2008, Harvey_2015}. In the present work, however, we consider spherically-symmetric halos and do not make use of offsets: we instead exploit the sensitivity of SL to the \emph{total} projected mass, while keeping in mind that this same sensitivity makes it difficult to disentangle the different mass components, a degeneracy that multi-wavelength approaches aim to break \citep{Beauchesne24, Massey_2010}.

It is essential to distinguish SI-SFDM, as studied here, from SIDM. Although both models lead to cored density profiles, they differ in fundamental ways. In SI-SFDM, the cores emerge from the self-interacting pressure of the scalar field, while in SIDM, they result from particle-particle scattering interactions \citep{Hui2017, Tulin_2018}. In \hyperref[app:si-sfdm-sidm]{Appendix \ref{app:si-sfdm-sidm}}, additional brief details on the differences between these two models are presented for clarification.

This study explores the potential of SL to constrain SI-SFDM models through their predicted soliton properties. The deflection angle in SL depends on both the soliton mass and the soliton radius. In the Thomas--Fermi regime the soliton radius is fixed by the microphysics (the scalar field mass and self-interaction scale), whereas its mass is not fixed by these parameters but is a bounded free parameter set by the formation history of the halo (see Sec.~\ref{subsec:microphysics}). By confronting theoretical predictions with SL observations, we establish constraints on these soliton characteristics, thereby probing the viability of SI-SFDM as an alternative to standard dark matter paradigms.

The key SL observable is the location of multiple images created by the lensing deflector, which enables inference of deflection angles. We assess whether SL observations can distinguish SI-SFDM models from the canonical Navarro-Frenk-White (NFW) density profile \citep{Navarro_1997}, the widely adopted dark matter density distribution for the CDM framework. Our analysis focuses on identifying deflection-angle discrepancies exceeding $2''$ relative to the NFW prediction. This threshold is deliberately conservative: parametric strong-lens models of galaxy clusters reproduce the observed multiple-image positions with a root-mean-square residual of typically $0.3$--$0.5''$ \citep{Limousin_2024}, well above the astrometric precision on individual image positions ($\lesssim0.1''$). A deviation of $2''$ therefore exceeds the systematic modelling accuracy by several times and constitutes a robustly distinguishable signal, whereas sub-arcsecond differences are comparable to the modelling residuals and cannot be reliably attributed to a soliton. This methodology provides a framework for deriving bounds on soliton parameters and for quantifying deviations of SI-SFDM models from the NFW paradigm.

The structure of this paper is as follows: \hyperref[sec:self-interacting model]{Section \ref{sec:self-interacting model}} introduces the self-interacting scalar field model. \hyperref[sec:gravitational-lensing]{Section \ref{sec:gravitational-lensing}} presents the fundamental equations for calculating deflection angles in strong gravitational lensing. \hyperref[sec:dm-halos]{Section \ref{sec:dm-halos}} explains the methodology for modeling dark matter halo profiles and constructing the corresponding density distributions. In \hyperref[sec:results]{Section \ref{sec:results}}, we present the lensing estimators applied to self-interacting dark matter halos. \hyperref[sec:comparison]{Section \ref{sec:comparison}} compares our results with previous studies to validate and contextualize our findings. Finally, \hyperref[sec:conclusion]{Section \ref{sec:conclusion}} summarizes the key conclusions of this work.

\section{Self-interacting scalar field dark matter} \label{sec:self-interacting model}

\subsection{The scalar field Lagrangian}
We consider the following Lagrangian to describe the real scalar field dark matter $\phi$,
\be 
\mathcal{L}_{\phi} = -\frac{1}{2}g^{\mu\nu}\partial_{\mu}\phi\partial_{\nu}\phi - V(\phi),\label{eq:lagrangian-sfdm}
\ee
where $g^{\mu\nu}$ is the inverse metric, the first term is the standard kinetic term and $V(\phi)$ is the potential given by, 
\be
V(\phi) = \frac{m^2}{2}\phi^2 + V_I(\phi), \label{eq:potential-v-sfdm}
\ee
where $V_I(\phi)$ is the self-interaction potential and here we work in natural units, $c = \hbar = 1$. We express the scalar field potential (\ref{eq:potential-v-sfdm}) as the sum of a dominant quadratic term and a secondary quartic  self-interacting potential:
\be 
V_I(\phi) = \frac{\lambda_4}{4}\phi^4. \label{eq:v_i-self-interacting}
\ee
where $\lambda_4$ is the self-interaction strength and takes in this study the positive sign.
In the weak gravity regime and neglecting the Hubble expansion, the scalar field follows a nonlinear Klein-Gordon equation,
\be
\ddot{\phi} -\nabla^2\phi+m^2(1+2\Phi_{\rm N})\phi  +(1+2\Phi_{\rm N})\frac{dV_I}{d\phi}=0 
\label{eq:phi-without-expansion-sfdm}.
\ee
This equation has been derived by applying the principle of least action with the Lagrangian (\ref{eq:lagrangian-sfdm}) and $\Phi_{\rm N}$ is the gravitational potential.
In the non-relativistic regime, relevant for astrophysical and large-scale structures,  it is useful to introduce a complex scalar field $\psi$ by \citep{Hu2000,Hui2017},
\be
\phi= \frac{1}{\sqrt{2m}} ( \psi e^{-imt} +  \psi^* e^{imt}) ,
\label{eq:phi-psi}
\ee
This allows us to separate the fast oscillations at frequency $m$ from the slower dynamics described by $\psi$ that follow the evolution of the density field and of the gravitational potential. Because $\psi$ varies on timescales much longer than $1/m$, its second-order time derivative is negligible compared with the first, $|\ddot\psi|\ll m|\dot\psi|$ (the slowly-varying-envelope, non-relativistic approximation). Substituting the decomposition (\ref{eq:phi-psi}) into Eq.(\ref{eq:phi-without-expansion-sfdm}) and dropping this term reduces the Klein--Gordon equation to the Schrödinger equation (\ref{eq:Schrodinger-real-1_sfdm}).
\be
i \frac{\partial\psi}{\partial t} = -\frac{1}{2 m} \nabla^2 \psi + m( \Phi_{\rm N} +\Phi_{\rm I})\, \psi,
\label{eq:Schrodinger-real-1_sfdm}
\ee
where we have defined the following self-interaction potential to make the Schr\"odinger equation (\ref{eq:phi-without-expansion-sfdm}) more user-friendly,
\begin{equation}
\Phi_{\rm I}(\rho) =\frac{d {\cal V}_I}{d\rho} ,
\label{eq:vi-phi-i-sfdm}
\end{equation}
where $\rho$ is the ultra-light scalar density,
\begin{equation}
\rho = m \psi \psi^*, \label{eq:rho-density-sfdm}
\end{equation}
and  $\mathcal{V_I} $  is the non-relativistic self-interacting scalar field potential resulting from the replacement of the decomposition (\ref{eq:phi-psi}) in the definition of $V_I$ (\ref{eq:v_i-self-interacting}). 

It is important to note that equation (\ref{eq:Schrodinger-real-1_sfdm}) resembles a Gross-Pitaevskii equation, with the key difference that the Newtonian potential \(\Phi_N\) is not externally imposed but instead originates from the self-gravity of the scalar field. Consequently, the Poisson equation governing the gravitational potential is given by:
\be
\nabla^2 \Phi_N = 4 \pi {\cal G}_N \rho.
\ee
\subsection{Madelung transformation}

Simple configurations can be understood from the hydrodynamical picture that follows from
the Madel\"ung transform \citep{Madelung1926}
\be 
\psi= \sqrt{\frac{\rho}{m}} e^{iS} , \;\;\; \mbox{where} \;\; \rho = m |\psi|^2 , 
\ee
where the dark matter velocity is identified as $\vec v = \nabla S/m$. The real and imaginary parts of
the Schr\"odinger equation give the continuity and Euler equations
\bea
&& \partial_t \rho + \nabla \cdot (\rho \vec v ) = 0 , \nonumber \\
&& \partial_t \vec v + (\vec v \cdot \nabla) \vec v= - \nabla (\Phi_{\rm N} + \Phi_{\rm I} + \Phi_{\rm Q}) ,
\label{eq:Euler-1}
\eea
with 
\be 
\Phi_{\rm I}= \frac{\rho}{\rho_a} \doteq \rho \frac{3 \lambda_4}{4 m^4} , \;\;\; \Phi_{\rm Q} =  - \frac {\nabla^2 \sqrt \rho}{2m^2 \sqrt \rho} ,
\label{eq:PhiI-PhiQ}
\ee
where $\Phi_{\rm Q}$ is the so-called quantum pressure.

\subsection{Hydrostatic equilibrium and Thomas-Fermi limit} \label{soliton}

As seen from Eq.(\ref{eq:Euler-1}), such scalar field models admit hydrostatic equilibria 
given by $\vec v=0$ and $\Phi_{\rm N}+\Phi_{\rm I}+\Phi_{\rm Q} = {\rm constant}$.
The spherically symmetric ground state is also called a soliton or boson star \citep{Seidel1994,Chavanis:2011zi,Chavanis:2011zm,Harko:2011jy,Brax:2019fzb}. 

In the Thomas-Fermi regime that we will consider in this paper, this soliton
is governed by the balance between gravity and the repulsive force associated with the 
self-interactions (for $\lambda_4 > 0$). This means that $\Phi_{\rm Q} \ll \Phi_{\rm I}$ over most of the
extent of the soliton and the Laplacian term $- \frac{\nabla^2\psi}{2m}$ can be neglected
in Eq.(\ref{eq:Schrodinger-real-1_sfdm}). Then, the wavefunction reads 
$\psi(r,t) = e^{-iE t} \hat\psi(r)$ with
\be 
\Phi_{\rm N} +\Phi_{\rm I}=\frac{E}{m} .
\label{eq:bal-TF}
\ee
The soliton density profile is given by \cite{Chavanis:2011zi,Harko:2011jy,Brax:2019fzb}

\be
\rho_{\rm sol}(r) = \rho_{0{\rm sol}} \frac{\sin(\pi r/R_{\rm sol})}{\pi r/R_{\rm sol}} , 
\label{eq:rho-soliton-1}
\ee
with the radius \citep{PhysRevD.109.043516, Brax:2019fzb}
\be
R_{\rm sol} = \pi r_a , \;\; \mbox{with} \;\; r_a^2 = \frac{3 \lambda_4}{16\pi {\cal G}_N m^4}
= \frac{1}{4\pi {\cal G}_N \rho_a} .
\label{eq:Rsol-1}
\ee
In fact, outside of the radius $r_a$ where Eq.(\ref{eq:rho-soliton-1}) would give
a zero density we can no longer neglect $\Phi_{\rm Q}$ and the exact solution develops
an exponential tail at large radii.
Nevertheless, from Eq.(\ref{eq:PhiI-PhiQ}) we can see that the approximation
(\ref{eq:rho-soliton-1}) is valid up to $r \lesssim R_{\rm sol}$ for
\be
\Phi_{\rm Q} \ll \Phi_{\rm I} : \;\;\; \frac{\rho_{0\rm sol}}{\rho_a} \gg \frac{1}{r_a^2 m^2} .
\label{eq:TF-1}
\ee

\subsection{Soliton parameters and the underlying microphysics}\label{subsec:microphysics}

The soliton is characterized by two independent quantities, fixed by physically distinct mechanisms. Its radius is set by the microphysics: from Eq.(\ref{eq:Rsol-1}), $R_{\rm sol}=\pi r_a$ with $r_a^2 = 3\lambda_4/(16\pi{\cal G}_N m^4)$, so it depends only on the combination $\lambda_4/m^4$. Its mass, equivalently $\rho_{0\rm sol}$, is not fixed by the microphysics; in the Thomas--Fermi regime it is a free parameter of the soliton family, set by the formation history and parametrized here by $\alpha$ (Sec.~\ref{sec:dm-halos}).

A lensing measurement of $R_{\rm sol}$ therefore constrains only the combination $\lambda_4/m^4$ and cannot separate $m$ and $\lambda_4$. An independent relation is provided by the self-interaction cross section per unit mass \citep{Brax:2019fzb},
\be
\frac{\sigma}{m} = \frac{9\lambda_4^2}{8\pi m^3} ,
\label{eq:cross-section}
\ee
which depends on a different combination of the two parameters. Eliminating $\lambda_4$ with Eq.(\ref{eq:Rsol-1}) yields
\be
\frac{\sigma}{m} = 32\pi {\cal G}_N^2\, m^5\, r_a^4 ,
\label{eq:cross-section-combined}
\ee
so that, at fixed $r_a$, the self-interaction bound from cluster mergers, $\sigma/m \lesssim 1~{\rm cm^2/g}$ \citep{Brax:2019fzb}, translates into an upper bound on $m$ and, through Eq.(\ref{eq:Rsol-1}), on $\lambda_4$. We emphasize that lensing alone does not measure $m$ or $\lambda_4$: the bounds are indirect, and follow only once the lensing constraint on $r_a$ is combined with the cross-section bound. Table~\ref{tab:params} lists these quantities for the two soliton sizes considered in this work. The resulting upper bounds, $m \lesssim {\rm few}\times10^{-5}$~eV and correspondingly $\lambda_4 \sim 10^{-18}$, lie within the range allowed for this model \citep{Brax:2019fzb}. Because $\sigma/m \propto m^5$, this conclusion is robust to the precise value adopted for the bound: even a factor-of-two tighter constraint lowers $m_{\rm max}$ by only $\sim15\%$.

\begin{table*}[t]
\centering
\begin{tabular}{cccccc}
\toprule
$r_a$ & $R_{\rm sol}$ & $\lambda_4/m^4$ & $m_{\rm max}$ & $\lambda_4$ & $\sigma/m$ \\
$[{\rm kpc}]$ & $[{\rm kpc}]$ & $[{\rm eV^{-4}}]$ & $[{\rm eV}]$ & & $[{\rm cm^2/g}]$ \\
\midrule
$5$  & $15.7$ & $6.9\times10^{-2}$ & $7.7\times10^{-5}$ & $2.4\times10^{-18}$ & $1.0$ \\
$15$ & $47.1$ & $6.2\times10^{-1}$ & $3.2\times10^{-5}$ & $6.5\times10^{-19}$ & $1.0$ \\
\bottomrule
\end{tabular}
\caption{Microphysical parameters of the two soliton configurations. The radius $R_{\rm sol}=\pi r_a$ and the combination $\lambda_4/m^4$ are fixed by $r_a$ [Eq.(\ref{eq:Rsol-1})]. Imposing the cluster-merger bound $\sigma/m \lesssim 1~{\rm cm^2/g}$ [Eq.(\ref{eq:cross-section-combined})] gives the upper bound $m_{\rm max}$ on the particle mass and the corresponding $\lambda_4$; the last column shows $\sigma/m$ evaluated at $m_{\rm max}$.}
\label{tab:params}
\end{table*}

\section{Gravitational lensing formalism}\label{sec:gravitational-lensing}

Gravitational lensing is a phenomenon where light rays are bent as they pass near a massive object due to the curvature of spacetime, as described by General Relativity \citep{Einstein:1936,Schneider:1992}.
This bending of light is characterized by two related quantities: the deflection angle, which is directly tied to the observed positions of the multiple images, and the excess surface mass density, a quantity derived from the projected mass and commonly used in weak lensing. Both follow from the projected mass distribution of the lens, and are the quantities we compute in this work.

\subsection{Deflection angle}
The deflection angle $\alpha$ describes how light is bent as it passes near a massive object. In the lensing process, the angular positions of the source $\beta$, the image $\theta$, and the deflection angle are related by the lens equation \citep{Einstein:1936,Schneider:1992,Narayan:1997,Meneghetti:2021},
\be 
\vec\beta = \vec\theta - \frac{D_{LS}}{D_{S}}\hat{\vec\alpha}(\vec\xi), \label{eq:lensing-eq}
\ee
where $D_{LS}$ is the angular diameter distance between the lens and the source, and $D_S$ is that between the observer and the source. The vector $\vec\xi=\vec\theta\,D_L$ is the impact parameter in the lens plane, with $D_L$ the observer--lens angular diameter distance, so that the deflection angle can equivalently be regarded as a function of $\vec\theta$. Defining the reduced deflection angle as follows,
\be 
\vec\alpha(\vec\theta) = \frac{D_{LS}}{D_{S}}\hat{\vec\alpha}(\vec\theta), \label{eq:deflection-angle-3}
\ee
the lens equation now reads:
\be 
\vec\beta = \vec\theta - \vec\alpha(\vec\theta). \label{eq:lensing-eq-2}
\ee
This angle is critical in strong gravitational lensing, where it explains phenomena such as multiple images and Einstein rings.
In many cases, such as gravitational lensing by galaxy clusters, the physical size of the lens is typically much smaller than the distances separating the observer, the lens, and the source. Consequently, the deflection of light occurs over a relatively short segment of its path. This allows us to adopt the thin-screen approximation, in which the lens mass is projected onto a single plane perpendicular to the line of sight, the lens plane. Within this approximation, the lensing matter distribution is effectively characterized by its surface density \citep{Bartelmann_2001},
\be
\Sigma(\vec{\xi})=\int\rho(\vec{\xi},z)dz, \label{eq:surface-mass}
\ee
where $\vec{\xi}$ is the impact parameter defined above and $\rho$ represents the three-dimensional density of the halo.

As long as the thin screen approximation holds, the total deflection angle can be determined by summing the contributions of all mass elements $\Sigma(\vec{\xi})d^2\xi$,
\be
\vec{\hat{\alpha}}(\vec{\xi}) =\frac{4 G}{c^2}\int\frac{(\vec{\xi}-\vec{\xi'})\Sigma(\vec{\xi'})}{\vert\vec{\xi}-\vec{\xi'}\vert^2}d^2\xi' . \label{eq:deflection-angle-1}
\ee
This angle is computed based on the mass distribution of the lens and the geometry of the source-lens-observer system.
It is a two-dimensional vector, but for axially symmetric lenses that we consider in this paper, it can be calculated in just one dimension. Therefore, for a symmetric mass distribution that we model, we have $\Sigma(\vec\xi) = \Sigma(|\vec\xi|)$.

\subsection{Excess of surface mass density}

The excess of surface mass density is a key quantity in gravitational lensing, particularly in weak lensing studies. It quantifies the difference between the average surface mass density within a given radius and the local surface mass density at that radius. Mathematically, it is defined as:
\be
\Delta \Sigma = \overline{\Sigma}(<R) - \Sigma(R),
\ee
where $\Sigma$ represents the projected surface mass density (\ref{eq:surface-mass}) and $\overline{\Sigma}$ is the average surface mass density within a circle of radius $R$ is given by:
\be
\overline{\Sigma}(<R) = \frac{2}{R^2} \int_{0}^{R} dR' \, R' \Sigma(R').
\ee
This magnitude is particularly useful for characterizing the mass distribution in galaxy clusters, where weak lensing effects are observed. Additionally, it is also related to the tangential shear, establishing a direct link between the observed lensing distortions and the mass distribution of the lens.

Together with the deflection angle, the excess of surface mass density forms a crucial pair of tools for understanding the mass distribution of lensing objects, such as galaxy clusters and dark matter halos, which are the focus of investigation in this work.

\section{Modeling the composite soliton–NFW halo}\label{sec:dm-halos}

Following the Thomas--Fermi picture introduced in Sec.~\ref{sec:intro}, in which solitonic cores are supported by repulsive self-interactions against gravity \citep{Chavanis:2011zi,Chavanis17c}, we adopt as a working hypothesis that, within halos formed through gravitational instability, such cores form dynamically and survive subsequent mergers as persistent configurations, while the outer halo retains the NFW profile of collisionless cold dark matter \citep{Schwabe2016}. This picture builds on the characterization of Thomas--Fermi solitons and halos in self-interacting scalar field dark matter of \cite{PhysRevD.109.043516}, and is supported by one- and three-dimensional simulations of soliton formation and mergers in the self-interacting regime \citep{Painter2026}, although a complete characterization of soliton formation at the high velocity dispersions of cluster cores remains an open question that we do not address here.

For late-time cosmological halos such as galaxy clusters, the self-interaction length scale $r_a$ is much smaller than the virial radius. Self-interactions therefore dominate only in the central region, where they support a solitonic core, while at larger radii the density converges to the NFW form, sustained by the velocity dispersion of the dark matter rather than by self-interaction forces. We accordingly model the halo as a composite profile: a Thomas--Fermi soliton within a transition radius $r_t$, joined to an NFW profile outside,
\bea
r < r_t: &&  \rho (r) = \rho_{0{\rm sol}} \frac{\sin(\pi r/R_{\rm sol})}{\pi r/R_{\rm sol}} , \nonumber \\
r_t < r < r_{200}: && \rho(r) =  \frac{\rho_s}{\frac{r}{r_s}\left(1+\frac{r}{r_s}\right)^2} ,
\label{eq:total-density}
\eea
where $\rho_{0{\rm sol}}$ and $R_{\rm sol}=\pi r_a$ are the central density and radius of the soliton, and $\rho_s$ and $r_s$ are the NFW scale density and radius; the outer profile recovers the expected $\rho \sim r^{-3}$ behaviour at large distances. In the remainder of this section we first fix the NFW halo (Sec.~\ref{method}) and then determine the soliton and its matching to the halo (Sec.~\ref{subsec:soliton-core}).

\subsection{The NFW halo}\label{method}
We first construct the NFW profile of the cluster that acts as the lens. The halo is defined by its total mass $M_{200}$, the mass enclosed within the radius $r_{200}$ at which the mean density is $200$ times the critical density of the Universe. Fixing the lens and source redshifts to $z_l = 0.2$ and $z_s = 1$, appropriate for cluster-scale lenses, we obtain $r_{200}$ from
\be 
r_{200} = \left(\frac{3M_{200}}{4\pi\,200\,\rho_{\rm crit}(z_l)}\right)^{1/3}. \label{eq:R200}
\ee
We then take the concentration $c_{200}$ at redshift $z_l$ from \citep{uchuu} and obtain the scale radius from its definition,
\be 
c_{200} = \frac{r_{200}}{r_s}, \label{eq:c}
\ee
and fix the scale density $\rho_s$ by imposing that the enclosed mass at $r_{200}$ equal the selected halo mass,
\be 
M_{\rm NFW}(r_{200}) = \int_0^{r_{200}} 4\pi r'^{2}\rho_{\rm NFW}(r')\,dr' = M_{200}. \label{eq:M200}
\ee
The resulting NFW parameters for the two halo masses considered in this work are listed in Table~\ref{tab:nfw-halos}.

\begin{table*}[t]
\begin{center}
\begin{tabular}{ccccc}
\toprule
 $M_{200}\,(\rm M_\odot)$ & $r_{200}$ (kpc) & $c_{200}$ & $\rho_s$ ($\rm 10^{5}\,M_\odot/kpc^3$) & $r_s$ (kpc)  \\
\midrule
$2\cdot 10^{15}$ & $2433.33$ & $3.5 \pm 1$ & $6.52 \pm 4.04$ & $695.24 \pm 198.64$ \\
$2\cdot 10^{14}$ & $1129.45$ & $4.5\pm 1$ & $11.35 \pm 5.70$ &  $250.99 \pm 55.77$ \\
\bottomrule
\end{tabular}
\caption{Parameters of the NFW profile for the two dark matter halos considered in this work. $M_{200}$ is the total mass within $r_{200}$, $c_{200}$ is the concentration, and $\rho_s$ and $r_s$ are the scale density and scale radius.}
\label{tab:nfw-halos}
\end{center}
\end{table*}

We adopt a conservative uncertainty $\Delta c_{200}=\pm1$ on the concentration, consistent with the halo-to-halo scatter reported in \citep{uchuu}, and propagate it to $\rho_s$ and $r_s$ using the expressions given in \hyperref[app:1]{Appendix~\ref{app:1}} [Eqs.(\ref{eq:delta_rhos}) and (\ref{eq:delta_rs})]. The uncertainty on $r_{200}$ is negligible, as it is fixed by the choice of $M_{200}$ through Eq.(\ref{eq:R200}) in a flat $\Lambda$CDM cosmology with $H_0 = 70\,\mathrm{km\,s^{-1}\,Mpc^{-1}}$ and $\Omega_M = 0.3$.

\subsection{The solitonic core and its matching to the halo}\label{subsec:soliton-core}
The soliton is joined to the NFW halo at the transition radius $r_t$, defined by
\be
M_{\rm sol}(r_t) = \alpha\,M_{\rm NFW}(r_t), \label{eq:mass-factor}
\ee
where $M_{\rm sol}(r_t)$ and $M_{\rm NFW}(r_t)$ are the soliton and NFW masses enclosed within $r_t$,
\bea
M_{\rm sol}(r_t) &=& \int_0^{r_t} 4\pi r'^{2}\rho_{\rm sol}(r')\,dr', \nonumber \\
M_{\rm NFW}(r_t) &=& \int_0^{r_t} 4\pi r'^{2}\rho_{\rm NFW}(r')\,dr'. \label{eq:mass-soliton}
\eea
The factor $\alpha$ measures how much soliton mass is enclosed within $r_t$ relative to the NFW value: $\alpha=1$ replaces the inner NFW by a soliton of equal enclosed mass (a mass-conserving swap), whereas $\alpha>1$ places additional mass in the core. As discussed in Sec.~\ref{subsec:microphysics}, the soliton mass is not fixed by the microphysics but is a bounded free parameter; we therefore treat $\alpha$ as free, restricted to the Newtonian regime and to configurations for which the total halo mass changes only marginally. There is, moreover, no universal scaling relation confirmed by simulations that would tie the soliton mass to the halo mass \citep{2022PhRvD.105l3528G}.

For given $r_a$ and $\alpha$, the transition radius $r_t$ and the soliton central density $\rho_{0{\rm sol}}$ are determined simultaneously by two conditions: the enclosed-mass matching of Eq.(\ref{eq:mass-factor}), which fixes $\rho_{0{\rm sol}}$ for a given $r_t$, and the continuity of the density at the transition, $\rho_{\rm sol}(r_t)=\rho_{\rm NFW}(r_t)$, which selects $r_t$. The density is therefore continuous across $r_t$, while the mass enclosed within $r_t$ is fixed by $\alpha$. The density derivative is not imposed to be continuous: for the most massive cores the soliton branch, which follows the Thomas--Fermi $\mathrm{sinc}$ profile and approaches its first zero near $R_{\rm sol}$, meets the NFW branch with a change of slope. This feature is smoothed in the exact solution by the exponential tail beyond $R_{\rm sol}$ (Sec.~\ref{soliton}) and, being a localized feature of the three-dimensional density, has a negligible effect on the projected, integrated lensing observables.

In the following we apply this construction to two halos of typical cluster scale, $M_h = 2\cdot10^{15}\,\mathrm{M_\odot}$ and $M_h = 2\cdot10^{14}\,\mathrm{M_\odot}$, and examine the resulting density and mass profiles as $\alpha$ and $r_a$ vary.

\subsubsection{Halo model of $M_h=2\cdot10^{15}M_{\odot}$}\label{subsec:15}

Here we present the results for a typical cluster-mass halo, $M_h=2\cdot10^{15}\,M_{\odot}$, later compared with Abell 2390. Table \ref{tab:solitons-15} outlines the soliton configurations for this halo, including key parameters such as the self-interacting scale $r_a$, inherently linked with the soliton core radius ($R_{sol} = \pi r_a$), the transition radius, $r_t$, the central density, $\rho_c$, the soliton mass, $M_{sol}$, the soliton's fractional contribution to the total system mass, $f_{sol}$, and its relative change in the total halo mass, $\Delta M_h \equiv [M_{\rm tot}(r_{200}) - M_{200}]/M_{200}$, where $M_{\rm tot}$ is the total enclosed mass of the composite soliton-plus-NFW profile.

\begin{table*}[ht]
\begin{center}
\begin{tabular}{ccccccc}
\toprule
\multicolumn{7}{c}{Halo $M = 2\cdot 10^{15}\,\rm M_\odot$} \\
\midrule
 $\alpha$ & $r_a$ (kpc) & $r_t$ (kpc) & $\rho_c$ ($\rm M_\odot/kpc^3$) & $M_{\rm sol}$ ($\rm M_\odot$) & $f_{\rm sol}\,(\%)$ & $\Delta M_h\,(\%)$ \\
\midrule
$1$ & $5$  & $10.90$ & $1.02\cdot 10^{8}$ & $3.31\cdot 10^{11}$ & $0.02$ & $0$ \\
$1$ & $15$ & $32.06$ & $3.26\cdot 10^{7}$ & $2.75\cdot 10^{12}$ & $0.14$ & $0$ \\
$3$ & $5$  & $14.50$ & $3.64\cdot 10^{8}$ & $1.75\cdot 10^{12}$ & $0.09$ & $0.06$ \\
$3$ & $15$ & $43.36$ & $1.14\cdot 10^{8}$ & $1.48\cdot 10^{13}$ & $0.74$ & $0.49$ \\
\bottomrule
\end{tabular}
\caption{Parameters of the solitonic profile for the halo of $M_h = 2 \cdot 10^{15} \rm M_\odot$, showing the central density, soliton mass, and fractional contribution to the system for different SI-SFDM models and $\alpha$ configurations.}
\label{tab:solitons-15}
\end{center}
\end{table*} 
\begin{figure*}[ht]
    \centering
    \includegraphics[height=7.cm,width=0.49\textwidth]{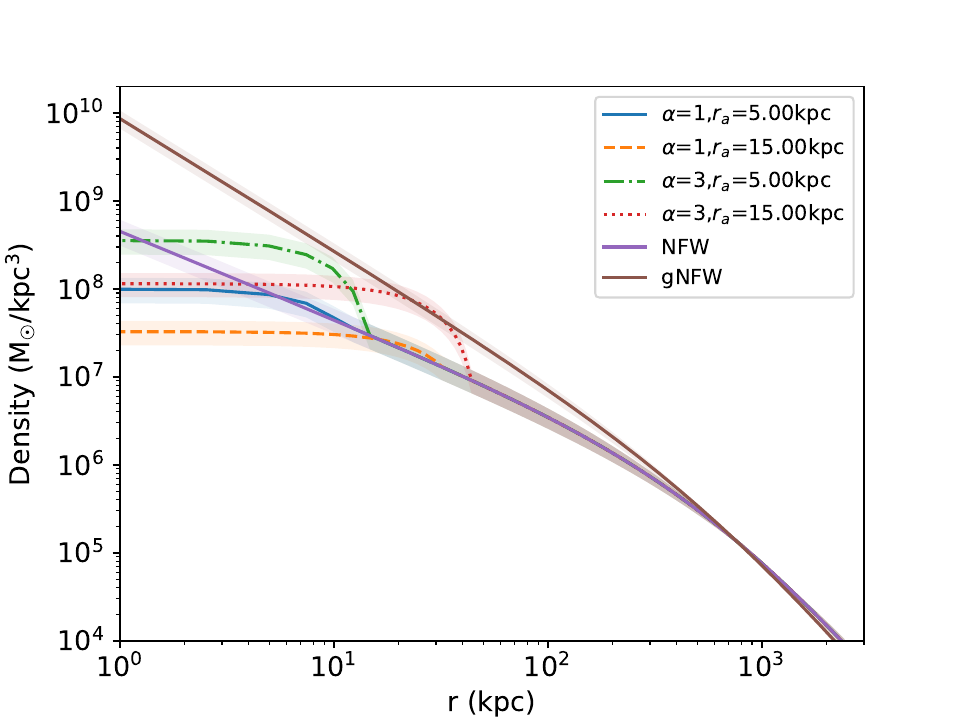}
    \includegraphics[height=7.cm,width=0.49\textwidth]{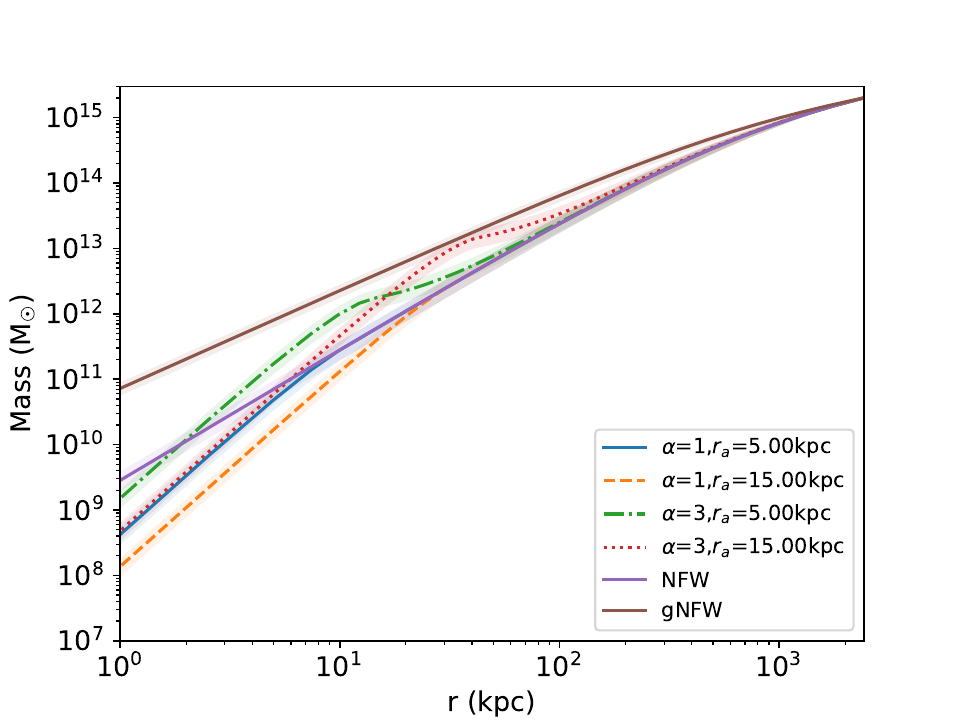}
    \caption{[$M_h=2\cdot10^{15}\rm M_\odot$] \textit{Left}: Density profile of the halo with $M_h = 2 \cdot 10^{15}\rm M_\odot$, showing the impact of the solitonic core at small scales and the recovery of the NFW behavior at larger radii. NFW profile in purple solid line for reference. \textit{Right}: Mass profile of the halo with $M_h = 2 \cdot 10^{15}\,\rm M_\odot$. The soliton's contribution is evident at small radii but remains negligible in the total mass at larger scales. NFW profile in purple solid line for reference.}
    \label{fig:density-15}
\end{figure*}
Figure \ref{fig:density-15} illustrates the density and mass profiles for these configurations. Alongside the NFW and soliton profiles we also show a generalized NFW (gNFW) profile. The gNFW is not a separate physical model but a one-parameter generalization of NFW with a steeper inner slope, constructed with the same concentration and the same total mass as the NFW, so that only the inner slope differs. It therefore represents, in a controlled way, the systematic uncertainty of the central density profile, whose validity in the innermost regions is debated because of baryonic processes and alternative dark matter models that can significantly influence the density near the core. We include it as a degeneracy test: it quantifies whether a steeper inner halo, at the same concentration and total mass, can reproduce the localized signature of a soliton. This is why the gNFW appears markedly different from the NFW in the inner regions---by construction---while remaining identical at large radii. Further details on this profile can be found in \hyperref[app:2]{Appendix \ref{app:2}}.

The soliton parameters follow two clear trends (Table~\ref{tab:solitons-15}). At fixed $r_a$, increasing $\alpha$ raises the central density, soliton mass and mass fraction; for $r_a = 5\,\mathrm{kpc}$, $\rho_c$ and $M_{\rm sol}$ grow by factors of $\sim3.5$ and $\sim5$ from $\alpha=1$ to $\alpha=3$. At fixed $\alpha$, increasing $r_a$ enlarges the transition radius $r_t$ and the soliton mass $M_{\rm sol}$ but lowers $\rho_c$, as the same mass is spread over a larger core. In all configurations the soliton is a small fraction of the halo, with $f_{\rm sol}\le0.74\%$ and $\Delta M_h\le0.49\%$. As shown in Fig.~\ref{fig:density-15}, the solitonic core dominates at small radii and joins smoothly onto the NFW profile, which sets the mass budget at large radii.

\subsubsection{Halo model of $M_h=2\cdot10^{14}\,M_{\odot}$}
\begin{table*}[ht]
\begin{center}
\begin{tabular}{ccccccc}
\toprule
\multicolumn{7}{c}{Halo $M = 2\cdot 10^{14}\,\rm M_\odot$} \\
\midrule
 $\alpha$ & $r_a$ (kpc) & $r_t$ (kpc) & $\rho_c$ ($\rm M_\odot/kpc^3$) & $M_{\rm sol}$ ($\rm M_\odot$) & $f_{\rm sol}\,(\%)$ & $\Delta M_h\,(\%)$ \\
\midrule
$1$ & $5$  & $11.21$ & $6.20\cdot 10^{7}$ & $2.12\cdot 10^{11}$ & $0.11$ & $0$ \\
$1$ & $15$ & $33.94$ & $1.80\cdot 10^{7}$ & $1.74\cdot 10^{12}$ & $0.87$ & $0$ \\
$3$ & $5$  & $14.50$ & $2.18\cdot 10^{8}$ & $1.05\cdot 10^{12}$ & $0.52$ & $0.34$ \\
$3$ & $15$ & $43.36$ & $6.30\cdot 10^{7}$ & $8.15\cdot 10^{12}$ & $4.08$ & $2.71$ \\
\bottomrule
\end{tabular}
\caption{Parameters of the solitonic profile for the halo of $M_h = 2 \cdot 10^{14} \rm M_\odot$, showing the central density, soliton mass, and fractional contribution to the system for different SI-SFDM models and $\alpha$ configurations.}
\label{tab:solitons-14}
\end{center}
\end{table*}
\begin{figure*}[ht]
    \centering
    \includegraphics[height=7.cm,width=0.49\textwidth]{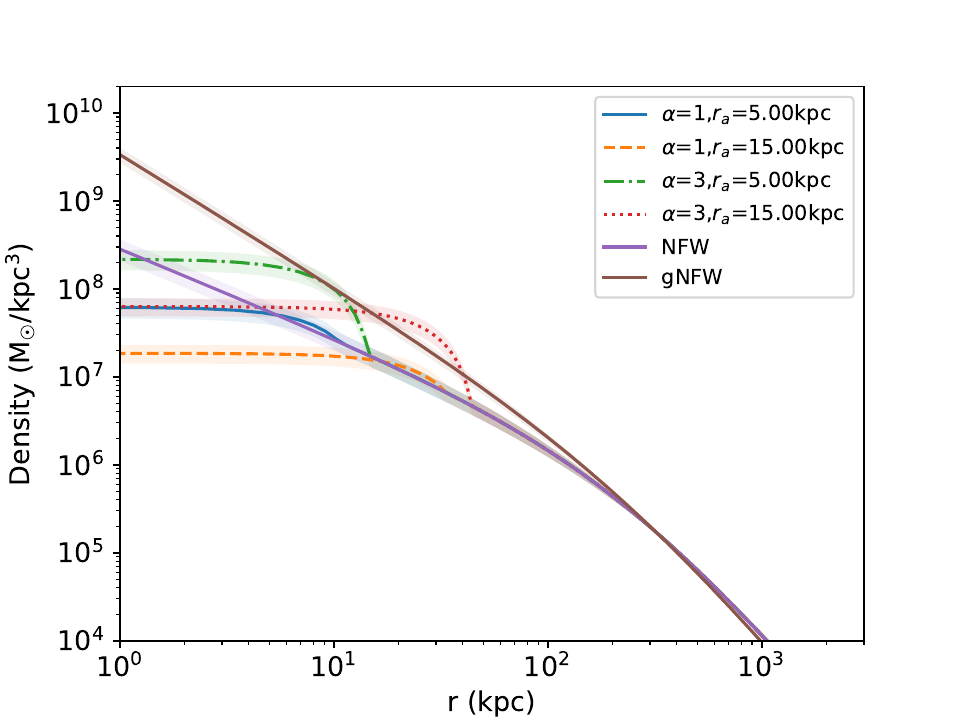}
    \includegraphics[height=7.cm,width=0.49\textwidth]{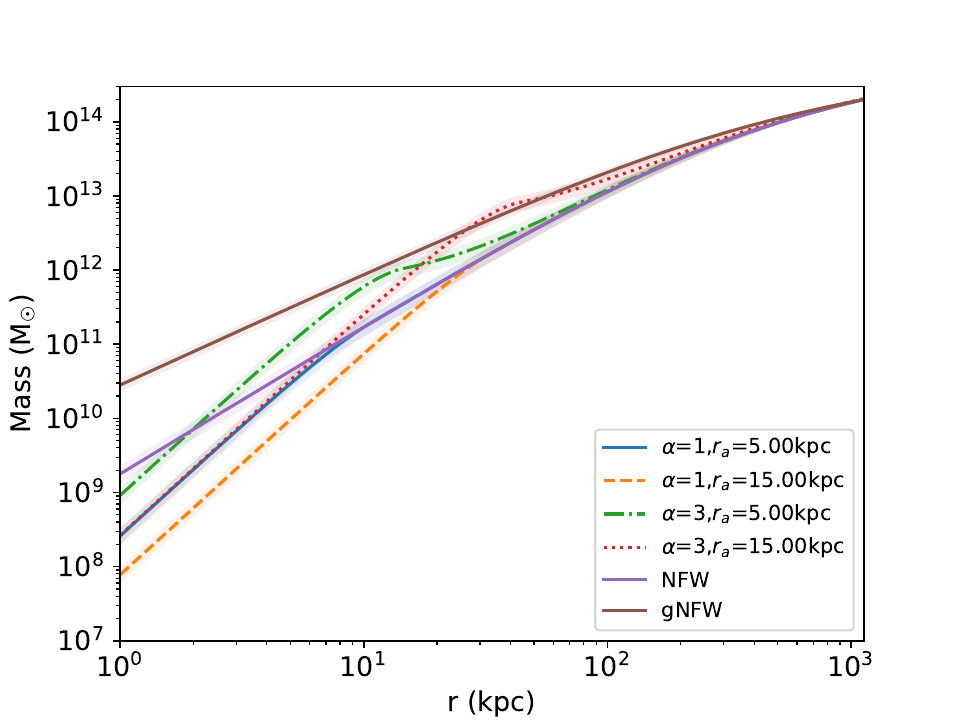}
    \caption{[$M_h=2\cdot10^{14}\rm M_\odot$] \textit{Left}: Density profile for the halo with $M_h = 2 \cdot 10^{14}\rm M_\odot$. The soliton's effect is more prominent in the central regions. NFW profile in purple solid line for reference. \textit{Right}: Mass profile for the halo with $M_h = 2 \cdot 10^{14}\rm M_\odot$. The soliton's contribution increases with $\alpha$, but the system retains the overall halo mass structure. NFW profile in purple solid line for reference.}
    \label{fig:density-14}
\end{figure*}
We now analyze the halo of mass $M_h = 2 \cdot 10^{14}\,M_{\odot}$, incorporating the self-interacting solitonic core. Table~\ref{tab:solitons-14} lists the soliton parameters, defined as in Table~\ref{tab:solitons-15}, and Fig.~\ref{fig:density-14} shows the corresponding density and mass profiles.

The lower-mass halo follows the same trends with $\alpha$ and $r_a$ as the $2\cdot10^{15}\,\mathrm{M_\odot}$ case. The key difference is that the soliton now represents a larger fraction of the halo: for $\alpha = 3$ and $r_a = 15\,\mathrm{kpc}$ it reaches $f_{\rm sol} = 4.08\%$ and $\Delta M_h = 2.71\%$, compared with $0.74\%$ and $0.49\%$ for the more massive halo. Soliton effects on the inner profile are therefore more pronounced at lower halo mass, while the total mass change stays below the $\sim3\%$ budget adopted in Sec.~\ref{subsec:soliton-core} even for the most extreme configuration. In both halos the composite profile displays the same dual-regime structure: a soliton-dominated core that joins smoothly onto the NFW profile, which dominates the mass at large radii.

\section{Results: lensing signatures of the soliton}\label{sec:results}
\begin{figure*}[ht]
\centering
\includegraphics[height=7.cm,width=0.49\textwidth]{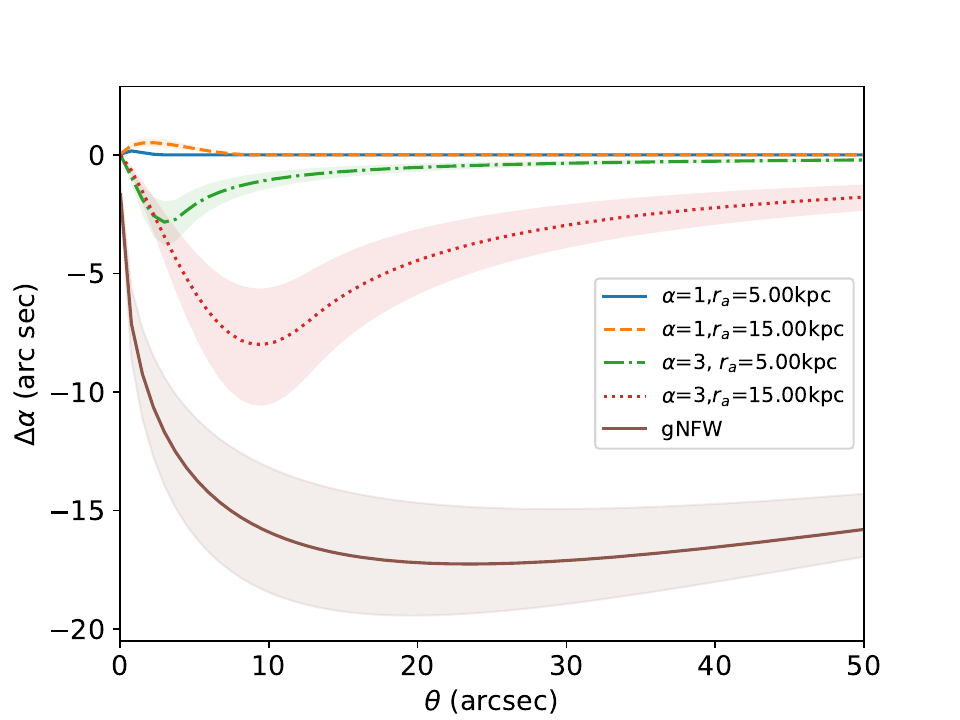}
\includegraphics[height=7.cm,width=0.49\textwidth]{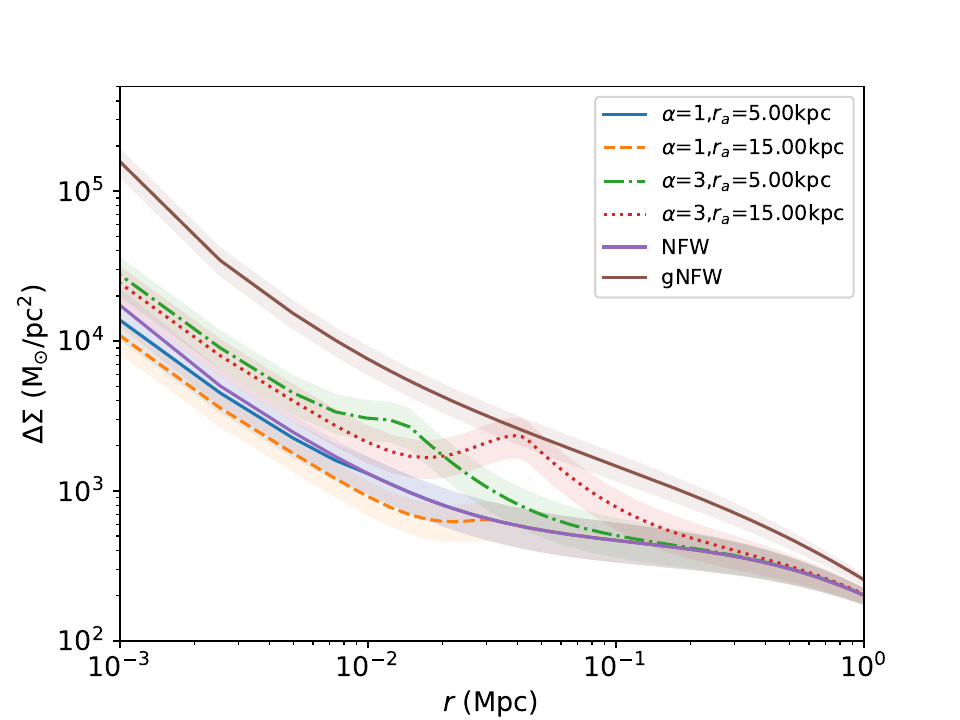}\\
\caption{[$M_h=2\cdot10^{15}M_{\odot}$] \textit{Left}: Difference in deflection angle ($\alpha_{\text{NFW}}- \alpha_{\text{sol}}$) relative to the NFW profile for different soliton configurations. The brown line represents the difference for the gNFW case $\alpha_{\text{NFW}}- \alpha_{\text{gNFW}}$. \textit{Right}: Excess of surface mass density. Brown line corresponds to the gNFW calculation.}
\label{fig:lensing-15}
\end{figure*}
\begin{figure*}[ht]
\centering
\includegraphics[height=7.cm,width=0.49\textwidth]{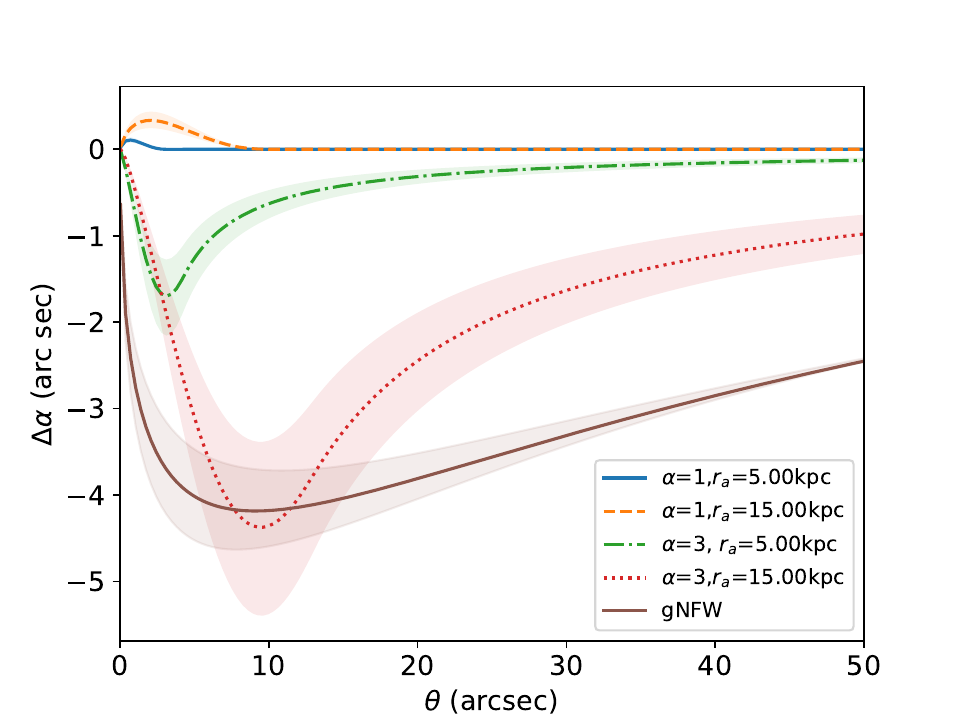}
\includegraphics[height=7.cm,width=0.49\textwidth]{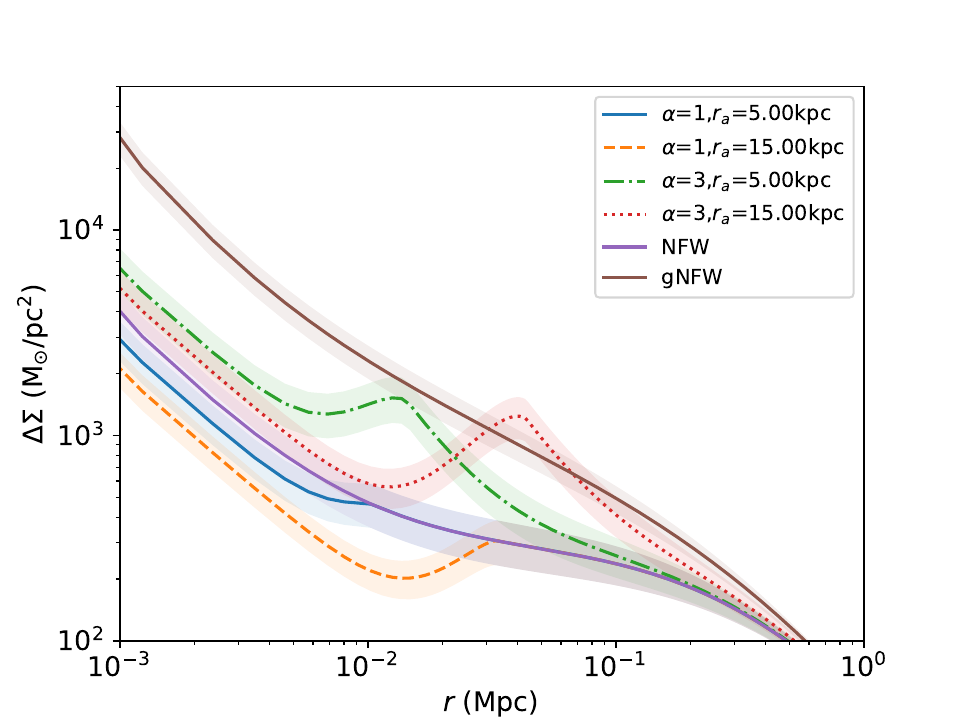}\\
\caption{[$M_h=2\cdot10^{14}\rm M_\odot$] \textit{Left}: Difference in deflection angle ($\alpha_{\text{NFW}}- \alpha_{\text{sol}}$) relative to the NFW profile for different soliton configurations. The brown line represents the difference for the gNFW case $\alpha_{\text{NFW}}- \alpha_{\text{gNFW}}$. \textit{Right}: Excess of surface mass density. Brown line corresponds to the gNFW calculation.}
\label{fig:lensing-14}
\end{figure*}

We present the calculations of the lensing estimators, focusing on deflection-angle differences between the widely used NFW profile (our baseline model) and the solitonic profile. We also quantify the excess surface mass density, a robust estimator on large scales, and compare the results to the gNFW profile.

For the massive halo ($M_h = 2 \times 10^{15}\,\mathrm{M_\odot}$), solitonic cores with three times the NFW-equivalent mass ($\alpha=3$) produce the largest deflection-angle differences (Fig.~\ref{fig:lensing-15}). For $r_a = 5\,\mathrm{kpc}$ and $r_a = 15\,\mathrm{kpc}$, the difference exceeds $2''$ at projected radii of $10$ and $31\,\mathrm{kpc}$, respectively, i.e. above the conservative $2''$ threshold adopted in this work and therefore within reach of high-resolution strong-lensing campaigns. The lighter cores ($\alpha=1$) produce sub-arcsecond differences that fall below this threshold.

The same behaviour holds for the lower-mass halo ($M_h = 2 \times 10^{14}\,\mathrm{M_\odot}$): the $\alpha=3$ cores generate deflection differences above $2''$ at $10\,\mathrm{kpc}$ ($r_a = 5\,\mathrm{kpc}$) and $32\,\mathrm{kpc}$ ($r_a = 15\,\mathrm{kpc}$), while the $\alpha=1$ cores again remain sub-arcsecond.

A degeneracy nevertheless appears when the soliton signal is compared with the gNFW profile (brown curve, Fig.~\ref{fig:lensing-14}). For the $\alpha=3$, $r_a=15\,\mathrm{kpc}$ configuration, the soliton and gNFW deflection differences partly overlap, so a steeper baryon-contracted inner halo could reproduce part of the soliton signal. This inner-slope systematic, modelled here through the gNFW, is the dominant source of confusion for the soliton signature, and it is the main reason why our results are stated as upper bounds rather than as a detection: distinguishing a soliton from a steep gNFW at the same concentration and total mass requires independent information on the inner profile, such as the resolved density decomposition available for A2390 (Sec.~\ref{sec:comparison}).

The surface mass density excess further underscores this duality. While large-scale (\( \gtrsim 0.1 \, \mathrm{Mpc} \)) results align between soliton and NFW halos, inner regions (\( \lesssim 10 \, \mathrm{kpc} \)) show marked differences. Observational limitations pose challenges: surveys like \citep{McClintock_2018} typically resolve scales \(\geq 0.1 \, \mathrm{Mpc}\), where soliton contributions are diluted. We stress, moreover, that the \citep{McClintock_2018} profile is a stacked average over many clusters in a mass bin rather than an individual system; the comparison is therefore meaningful only in a statistical, large-scale sense, since the soliton signature---localized in the core of an individual halo---is washed out by the stacking. Detecting it requires resolved strong-lensing measurements of individual clusters rather than a stacked weak-lensing profile. High-resolution lensing data, such as those from the James Webb Space Telescope \citep{Rigby_2023}, would be needed to probe the soliton-driven excess in the core.

\section{Comparison with observations}\label{sec:comparison}
In this section we compare our SI-SFDM model with observations. We focus on the most massive halo in our analysis, $M_h=2 \cdot 10^{15}\,\rm M_\odot$, for which cluster-scale data are available, and select the well-characterized galaxy cluster A2390 at redshift $z = 0.229$, close to our model's lensing redshift $z_l =0.2$.

The halo mass and NFW parameters derived from our model (Table \ref{tab:comparison}) are consistent with the uncertainty ranges reported in \citep{Newman_2013_1}, which combines strong/weak lensing and stellar kinematic data. While \citep{Newman_2013} also provides X-ray-based constraints, these are excluded here to maintain consistency with our lensing-centric methodology.
\begin{table*}[t]
\begin{center}
\begin{tabular}{ccccc}
\toprule
Cluster & $r_s$ (kpc) & $c_{200}$ & $\log M_{200}/\rm M_\odot$ & $r_{200}$ (kpc) \\
\midrule
A2390 & $763^{+119}_{-107}$ & $3.24^{+0.35}_{-0.31}$ & $15.34^{+0.06}_{-0.07}$ & $2470^{+112}_{-123}$ \\
$2\cdot10^{15}\,\rm M_\odot$ & $695.24 \pm 198.64$ & $3.5 \pm 1$ & $15.30$ & $2433.33$ \\
\bottomrule
\end{tabular}
\caption{Comparison of NFW parameters for A2390 (from \citep{Newman_2013_1}) and our SI-SFDM model for $M_h=2\cdot10^{15}\rm M_\odot$. Uncertainties for A2390 reflect combined lensing and stellar dynamics.}
\label{tab:comparison}
\end{center}
\end{table*}
\vspace{-1cm}
\begin{figure}[h!]
    \centering
    \includegraphics[width=1.1\linewidth]{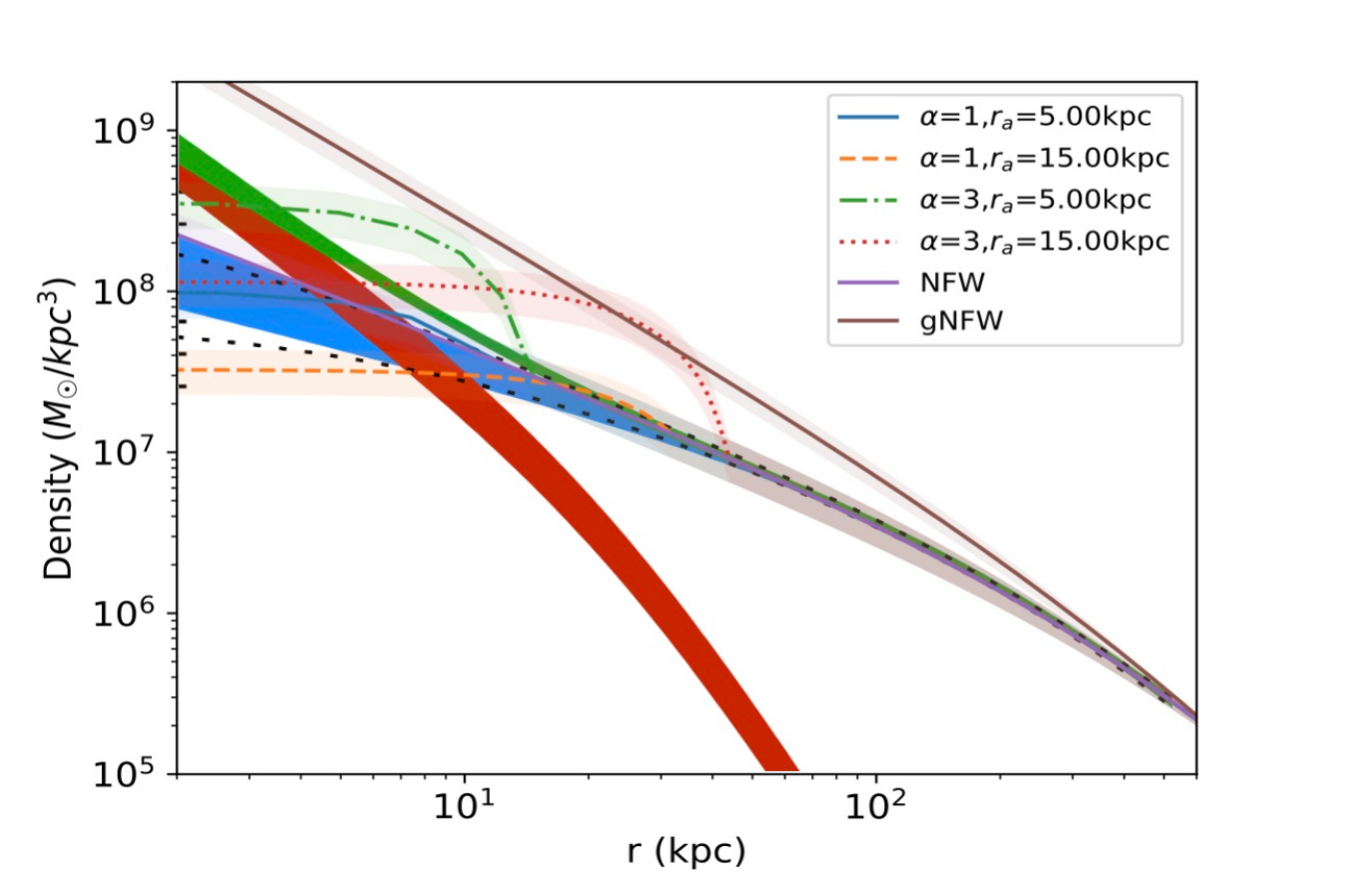}
    \caption{Comparison of the radial density profiles of the SI-SFDM dark matter models for the $M_h = 2 \cdot 10^{15} \rm M_\odot$ halo with A2390 observations and simulations from \citep{Newman_2013}. The blue, red, and green bands represent the radial density profiles of the DM halo, the stars in the BCG, and their total in A2390, respectively. The color scheme for the SI-SFDM profiles is consistent with that shown in Fig.~\ref{fig:density-15}. }\label{fig:comparison-density-15}
\end{figure}
\begin{figure*}[ht]
    \centering
    \includegraphics[width=0.7\linewidth]{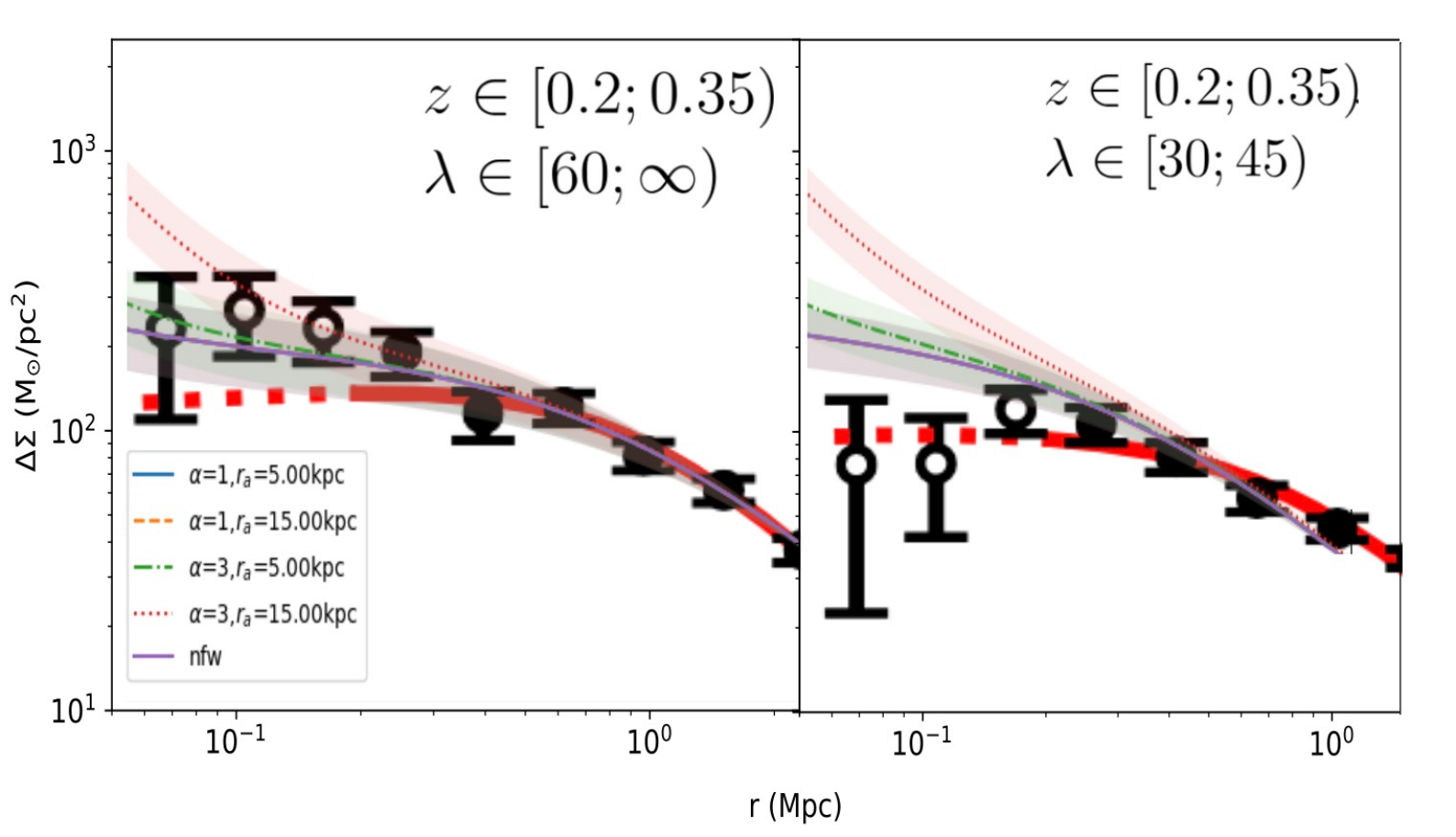}
    \caption{Comparison of excess surface mass density for the halos $M_h=2\cdot10^{15}\rm M_\odot$ (left) and $M_h=2\cdot10^{14}\rm M_\odot$ (right) with  \citep{McClintock_2018}. }
    \label{fig:comparison-excess}
\end{figure*}

Figure \ref{fig:comparison-density-15} compares the radial density profiles of our SI-SFDM models with A2390's profile. The density profile of A2390 includes contributions from dark matter (modeled with generalized NFW [gNFW; light blue] and classical NFW [cNFW; black dashed]), stars in the brightest cluster galaxy (BCG; red), and the total mass (green band).
 
These components are derived from observations of multiple images, lensing data, and kinematics. In the inner regions (\(r \lesssim 5\text{–}10\) kpc), where stars dominate, the total density profile steepens significantly. The stellar mass density in the model of A2390 reaches parity with DM at approximately 7 kpc, while enclosed mass equality occurs at around 12 kpc. Within 5 kpc, the DM fraction is about 25\%, increasing to 80\% within the three-dimensional half-light radius. 

We find that the NFW contribution in our model (purple) aligns closely with the corresponding component in A2390's profile (black dashed). In the inner regions, the enclosed dark matter mass of A2390 inferred from the Newman et al. NFW fit is $\simeq6\times10^{11}\,\mathrm{M_\odot}$ within $15\,\mathrm{kpc}$ and $\simeq2\times10^{12}\,\mathrm{M_\odot}$ within $30\,\mathrm{kpc}$, and the inner slope is measured to be shallower than NFW \citep{Newman_2013}. The massive cores ($\alpha=3$), which by construction enclose three times the NFW mass at $r_t$ ($M_{\rm sol}=1.75\times10^{12}\,\mathrm{M_\odot}$ for $r_a=5\,\mathrm{kpc}$ and $1.48\times10^{13}\,\mathrm{M_\odot}$ for $r_a=15\,\mathrm{kpc}$), exceed this budget by a factor of a few and overpredict the observed central density. We therefore adopt $M_{\rm sol}\lesssim10^{12}\,\mathrm{M_\odot}$ as an upper limit on the soliton mass for this cluster. The $\alpha=1$ configurations, for which the soliton and NFW enclosed masses are equal, remain consistent with the data, but their sub-arcsecond angular signatures place them below the current detection threshold.

We further compare our model's predictions for the halos studied in this work---at redshift $z_l = 0.2$ with masses $M_h = 2 \cdot 10^{15}\,\rm M_\odot$ and $M_h = 2 \cdot 10^{14}\,\rm M_\odot$---to the weak-lensing results of \citet{McClintock_2018} for the excess surface mass density $\Delta\Sigma$, shown in Figure~\ref{fig:comparison-excess}. \citet{McClintock_2018} constrain the mass--richness scaling relation of redMaPPer galaxy clusters in the Dark Energy Survey Year~1 data, deriving mean $\Delta\Sigma$ profiles for cluster subsets binned by photometric redshift and optical richness. Their analysis excludes scales $r<200$~kpc (open symbols, dashed lines) to avoid contamination from poorly constrained baryonic effects and miscentering.

In Figure~\ref{fig:comparison-excess}, the red line is the best-fit NFW model reported by \citet{McClintock_2018} for their stacked $\Delta\Sigma$ data, and our soliton+NFW predictions are overlaid with the color scheme of the earlier figures. The left panel corresponds to $M_h = 2 \cdot 10^{15}\,\mathrm{M_\odot}$ and the right panel to $M_h = 2 \cdot 10^{14}\,\mathrm{M_\odot}$, matching the figure caption. At large scales, our predictions for both halo masses agree with the data and with the \citet{McClintock_2018} model, confirming that the soliton has a negligible impact on the halo outskirts and that the total mass profile remains NFW-like there. In the inner regions the profiles differ; as stressed above, the stacked nature of the measurement precludes a direct detection of the soliton on those scales, since soliton-driven deviations are largest in the core, precisely where current weak-lensing data cannot resolve them. Testing the soliton predictions directly therefore requires resolving the cluster core: combining the resolution of JWST at $r<200\,\mathrm{kpc}$ with wide-field surveys such as the Rubin Observatory, and stellar-kinematic measurements able to probe the $\alpha=1$ regime. The agreement at larger radii confirms that the soliton+NFW model reproduces the established NFW lensing behaviour on the scales where baryonic uncertainties dominate.
\section{Conclusion}\label{sec:conclusion}

The differences in the gravitational-lensing signatures of cold dark matter and self-interacting scalar field dark matter provide an observational route to test these models. Using existing lensing data, we have shown that the SI-SFDM soliton parameters can be constrained at cluster scales. Our predictions agree with the weak-lensing mass calibration of \citet{McClintock_2018} at large scales, and the halo parameters of our $M_h=2\cdot10^{15}\,\mathrm{M_\odot}$ model are consistent with the multi-probe lensing and stellar-dynamics analysis of A2390 by \citet{Newman_2013_1}. From the inner density profile of A2390 we obtain an upper limit on the soliton mass, $M_{\rm sol}\lesssim10^{12}\,\mathrm{M_\odot}$: more massive cores ($\alpha=3$) overpredict the observed central dark matter density, while cores with $\alpha=1$ remain consistent but produce sub-arcsecond lensing signatures below the current detection threshold.

To strengthen these results, we propose a two-pronged approach for future work. First, we will incorporate solitonic profiles into analyses of observed multiple-image lensing systems, directly applying our model to real data. By initially neglecting error propagation, we aim to prioritize model discrimination and statistically robust comparisons, testing whether SI-SFDM can explain lensing features as a viable alternative to CDM. This approach will enable simulations of lensed images across SI-SFDM scenarios, probing whether input soliton parameters (e.g., \(\alpha\), \(r_a\)) can be accurately reconstructed from synthetic and observational data. Second, we will rigorously evaluate whether real data supports solitons as unique descriptors of halo cores or if degeneracies arise with other mass components. These efforts will focus on the central regions of galaxy clusters, where SI-SFDM deviations from CDM predictions are most pronounced, offering a direct test of scalar field dynamics on small scales.

Future facilities combining the resolution of Euclid and JWST with wide-field surveys such as the Rubin Observatory will extend this work, enabling tests of the SI-SFDM predictions at sub-arcsecond scales, in particular for the low-mass ($\alpha=1$) solitons that lie below current detectability.
These efforts will ultimately clarify whether self-interacting scalar fields represent a fundamental component of the dark sector or if new physics beyond current paradigms must be invoked.

\acknowledgments
This work received support from the French government under the France 2030 investment plan, as part of the Initiative d'Excellence d'Aix-Marseille Universit\'e -- A*MIDEX (AMX-21-RID-039).

\section*{Data Availability}
The data that support the findings of this article are not publicly available. The data are available from the authors upon reasonable request.

\appendix
%
%

\section{Self-interacting scalar field dark matter vs. self-interacting dark matter} \label{app:si-sfdm-sidm}
Although both SI-SFDM and SIDM produce cored density profiles, they do so through fundamentally different mechanisms, which we make explicit here. In SI-SFDM the dark matter is a coherent scalar field, and the core is supported by the effective pressure of its self-interaction, encoded in the potential term $\lambda_4\phi^4$ [Eq.(\ref{eq:v_i-self-interacting})]. The sign of $\lambda_4$ sets the character of the interaction: repulsive for $\lambda_4>0$, which stabilizes the soliton against gravitational collapse and is the case studied here, and attractive for $\lambda_4<0$. The core is therefore a static equilibrium configuration of a Bose--Einstein condensate, not the outcome of collisions.

In SIDM, by contrast, the dark matter is a gas of particles $\chi$ whose cores form through elastic scattering, typically mediated by a new force \citep{Tulin_2018}. In the scalar-mediator case the interaction Lagrangian reads
\begin{equation}
\mathcal{L}_{\text{SIDM}} = g_{\chi}\bar{\chi}\chi\phi,
\end{equation}
where $\phi$ is the mediator and $g_{\chi}$ the coupling. In the non-relativistic limit the interaction is described by a Yukawa potential,
\begin{equation}
V(r) = \pm\frac{\alpha_{\chi}}{r}e^{-m_{\phi}r},
\end{equation}
with dark fine-structure constant $\alpha_{\chi} = g_\chi^2/4\pi$ and mediator mass $m_\phi$. Here the sign is set by the mediator: a vector mediator gives an attractive potential ($-$) for $\chi\bar{\chi}$ scattering and a repulsive one ($+$) for $\chi\chi$ or $\bar{\chi}\bar{\chi}$, whereas a scalar mediator is purely attractive.

The essential difference is therefore one of mechanism: in SI-SFDM the core arises from the pressure of a coherent field and its properties follow from the potential $V(\phi)$, whereas in SIDM the core arises from repeated particle scattering and its properties follow from the velocity-dependent cross section. This distinction should be kept in mind when applying cross-section bounds derived for collisional SIDM, such as those from cluster mergers, to the scalar-field case considered here.

\section{Testing the consistency of NFW parameters with a steeper central density profile} \label{app:2}

The applicability of the NFW profile in the innermost regions of halos is debated, since observational tensions there are often attributed to baryonic effects such as active galactic nucleus (AGN) feedback and adiabatic contraction, which can steepen the central density. To assess whether such a steeper core could mimic the lensing signature of a soliton, we introduce a gNFW profile as a controlled, single-parameter variation of the inner slope,
\be 
\rho(r) = \frac{\rho_s}{\left(\frac{r}{r_s}\right)^\gamma\left(1+\left(\frac{r}{r_s}\right)^\eta\right)^{(\beta - \gamma)/\eta}} ,
\label{eq:gnfw}
\ee
where $\gamma$ is the inner slope, $\beta$ the outer slope, and $\eta$ controls the sharpness of the transition between them. We adopt a steeper inner slope than the NFW value, $\gamma = 1.5$ (versus $\gamma = 1$ for NFW), together with the standard outer slope $\beta = 3$ and $\eta = 1$. This choice, motivated by baryon-contracted profiles \citep{scheider15}, serves as a conservative estimate of a denser core in the absence of direct observational constraints on these scales.

To isolate the effect of the inner slope, we build the gNFW with the same halo mass $M_{200}$ and concentration $c_{200}$ as the NFW profile: $r_{200}$ and $r_s$ follow from Eqs.(\ref{eq:R200}) and (\ref{eq:c}), and the scale density $\rho_s$ is fixed by imposing the same total mass. The gNFW therefore differs from the NFW only in the inner slope, while sharing the global parameters that are most directly constrained by observations. The resulting parameters are listed in Table~\ref{tab:steep-profile}.
\begin{table}[h!]
\begin{center}
\begin{tabular}{cc}
\toprule
 $M_{200}\,(\rm M_\odot)$ & $\rho_s$ ($\rm 10^5\,M_\odot/kpc^3$)  \\
\midrule
$2\cdot 10^{15}$ & $4.71 \pm 4.06$ \\
$2\cdot 10^{14}$ & $8.50 \pm 5.69$ \\
\bottomrule
\end{tabular}
\caption{Scale density $\rho_s$ of the gNFW profile for the two halos. The remaining parameters ($r_{200}$, $c_{200}$ and $r_s$) are identical to the NFW values in Table~\ref{tab:nfw-halos}, since the gNFW is constructed with the same $M_{200}$ and $c_{200}$.}
\label{tab:steep-profile}
\end{center}
\end{table}

Because $M_{200}$ and $c_{200}$ are held fixed, a steeper inner slope is accommodated solely through the scale density $\rho_s$, without altering the global halo parameters or the large-radius profile. The gNFW thus provides a clean representation of the inner-slope systematic: in the main text (Sec.~\ref{sec:results}) we use it to assess to what extent a steep baryon-contracted core, at fixed concentration and total mass, can mimic the localized signature of a soliton.

\section{Uncertainty Propagation Expressions in NFW Density and gNFW Profile Parameters} \label{app:1}
In this appendix we present the expressions used for the error propagation in our calculations for the density profiles:
\be
\Delta r_s = \frac{r_{200}}{c^2}\Delta c \label{eq:delta_rs}
\ee
\begin{widetext}
NFW profile:
\be
\Delta \rho_s = \frac{
    M_{200} \left( 4 r_{200}^2 + 3 r_{200} r_s + 3 (r_{200} + r_s)^2 \ln\left( \frac{r_s}{r_{200} + r_s} \right) \right) 
}{
    4 \pi r_s^4 \left( r_{200} + (r_{200} + r_s) \ln\left( \frac{r_s}{r_{200} + r_s} \right) \right)^2
}\Delta r_s \label{eq:delta_rhos}
\ee
gNFW profile:
\be
\Delta \rho_s = \rho_s \Delta r_s \left| \frac{3}{2 r_s} + \frac{9 r_{200}}{10 r_s^2}\, \frac{{}_2F_1\!\left(\tfrac{5}{2}, \tfrac{5}{2}; \tfrac{7}{2}; -\tfrac{r_{200}}{r_s}\right)}{{}_2F_1\!\left(\tfrac{3}{2}, \tfrac{3}{2}; \tfrac{5}{2}; -\tfrac{r_{200}}{r_s}\right)} \right|
\ee
\end{widetext}

where \(\, _2F_1(a, b; c; z) \) is the hypergeometric function defined as the power series:  
\[
_2F_1(a, b; c; z) = \sum_{n=0}^{\infty} \frac{(a)_n (b)_n}{(c)_n} \frac{z^n}{n!},
\]
where \((q)_n\) is the Pochhammer symbol (rising factorial), given by:  
\[
(q)_n = q (q+1) (q+2) \cdots (q+n-1), \quad (q)_0 = 1.
\]

\newpage

\bibliography{june24}

\end{document}